\documentclass[
preprint
,showpacs
,showkeys
,nofootinbib
,aps
,amsfonts
,amssymb
,pdftex 
]{revtex4-1}

\usepackage{graphicx} 
\usepackage{amsmath, amssymb, bm} 
\usepackage[dvips]{color}

\usepackage{psfrag,epsfig}
\usepackage{natbib}
\usepackage{hyperref}

\def\){\right)} 
\def\({\left(} 
\def\]{\right]} 
\def\[{\left[}

\def\nLi{$^7\mathrm{Li}(n, \gamma)^8\mathrm{Li}$}

\def\P { \left(\frac{\stackrel{\rightarrow}{\nabla}}{M_c}-\frac{\stackrel{\leftarrow}\nabla}{M_n}\right)}

\def\erasepar#1{}

\begin{document}

\title{
Resonance Contribution to Radiative Neutron Capture on Lithium-7}

\author{%
Lakma Fernando $^a$}
\email{nkf22@msstate.edu}

\author{%
Renato Higa $^{b,c}$}
\email{higa@if.usp.br}

\author{%
Gautam Rupak $^a$}
\email{grupak@u.washington.edu}

\affiliation{$^a$ Department of Physics $\&$ Astronomy and 
HPC$^2$ Center for Computational Sciences, 
Mississippi State
University, Mississippi State, MS 39762, U.S.A. \\
$^b$ Kernfysisch Versneller Instituut, Theory Group, University 
of Groningen, 9747AA Groningen, The Netherlands \\
$^c$ Instituto de F\'\i sica, Universidade de S\~ao Paulo, 
C.P. 66318, 05314-970, S\~ao Paulo, SP, Brazil}

\begin{abstract}
Using halo effective field theory, we provide a model-independent calculation 
of the radiative neutron capture on lithium-7 over an energy range where the 
contribution from the $3^+$ resonance becomes important. 
One finds that a satisfactory description 
of the capture reaction, in the present single-particle approximation, 
suggests the use of a resonance width about three times larger than the 
experimental value. We also present power counting arguments that 
establish a hierarchy for electromagnetic one- and two-body currents. 
\end{abstract}


\maketitle

\section{Introduction}\label{sec_intro}

The field of nuclear structure and reactions have witnessed continuous 
progress and 
renewal of interest with the advent of present and future facilities that 
are able to provide high-intensity beams of very unstable, rare isotopes 
---so-called exotic nuclei--- whose physical properties are in the process 
of being uncovered. They populate areas in the nuclear chart far from the 
valley of stability, resembling more weakly-bound clusters rather than a 
tighter, shell-like structure. A subset in this exotic zoo comprises halo 
nuclei, systems formed by a tightly bound core nucleus surrounded by one or 
more loosely bound nucleons, with a slowly decreasing wave function tail 
that extends much farther than the effective core-nucleon interaction. 
Such extended and dilute configuration leads to threshold phenomena 
with consequences to low-energy nuclear 
astrophysics~\cite{Bertulani:2009mf,Rauscher:2010pu}. 

An example of astrophysically relevant halo nucleus is boron-8, with 
dominant configuration of a beryllium-7 core loosely holding a proton by 
about a tenth of MeV. This nucleus plays a decisive role in our 
current understanding of neutrino physics. Underground detectors like 
SNO and Super-K are mainly sensitive 
to neutrinos released from the $\beta$-decay of $^8$B in 
the Sun. The capture rate of a proton by beryllium-7 that produces boron-8, 
$^7{\rm Be}(p,\gamma)^8{\rm B}$, is crucial for determining the 
initial electron neutrino flux that eventually transmutes to other 
neutrino flavors on its way to detection. The reaction rate for 
conditions in the solar core sharply peaks around 20 keV, the Gamow peak, 
while experimental data remain above 100 keV. 
Measurements at lower energies are extremely 
difficult, mainly due to 
the Coulomb repulsion. The solar neutrino flux is therefore dependent 
on theoretical extrapolations of current data to lower energies. 
Available theoretical 
approaches~\cite{Bertulani1996293,Navratil:2005xm,*Navratil:2006tt,
Navratil:2010jn,Bennaceur:1999hv,Shul'gina1996197,*Grigorenko:1999mp,
Descouvemont1994341,*Descouvemont:2004hh,Tombrello:1965,PhysRevC.68.045802,
Huang:2008ye} 
that use $p$-$^7{\rm Be}$ data alone, even 
with scattering information, are not well-constraining. 
The mirror system $n$-$^7{\rm Li}$ then becomes an important ingredient for 
benchmarking purposes ---the Coulomb force between the initial particles is 
absent and more precise data at very low (sub-keV) energies are available. 
It also strongly 
constrains the $^7{\rm Be}(p,\gamma)^8{\rm B}$ if mirror symmetry, 
which ought to have its origins in the (accidental) isospin symmetry of QCD, 
is invoked. 
Besides, $n$-$^7{\rm Li}$ has its own astrophysical interest, bridging 
the path to the formation of heavier elements in inhomogeneous big-bang 
nucleosynthesis models~\cite{kawano:1991ApJ372}. 

The \nLi~ cross section was calculated recently in effective field theory 
(EFT) for halo nuclei at low energy~\cite{Rupak:2011nk}. In this framework, 
the tight $^7{\rm Li}$ core, inert and structureless at leading order (LO), 
the loosely bound neutron, and the external soft photons are the relevant 
degrees of freedom. The main assumption of the approach is a single-particle 
approximation somewhat similar to models like 
Refs.~\cite{Tombrello:1965,Typel:2004us,PhysRevC.68.045802,Huang:2008ye}, 
where the valence nucleon interacts with the core via a Woods-Saxon 
potential. In halo EFT, however, a systematic and model-independent 
expansion of observables is achieved through the use of an expansion 
parameter ---formed by the ratio of a soft scale $Q$, associated 
with the shallowness of the valence neutron, and a hard scale $\Lambda$, 
related to the tightness of the core. 
Moreover, the formalism guarantees unambiguous inclusion of electromagnetic 
interactions that preserve the required symmetry constraints, 
such as gauge invariance~\cite{Phillips:2010dt,*Hammer:2011ye,Rupak:2011nk}. 
In Ref.~\cite{Rupak:2011nk} only the leading E1 transition from initial 
$s$-wave continuum to the $p$-wave $^8{\rm Li}$ ground state was considered. 
The authors found the poorly known $p$-wave effective range to be the 
leading source of uncertainty. 
In this work we extend the previous one in several directions, namely, 
we include explicitly the capture to the first excited state of $^8{\rm Li}$, 
the contribution from the low-lying $p$-wave resonance at $\sim 0.22$ MeV, 
and elaborate on the power counting for the two-body currents. 
With these extra ingredients, one sets the formalism that paves the way to 
handle the more weakly bound $p$-$^7{\rm Be}$ mirror system. 

The paper is organized as follows. In Sec.~\ref{sec_theory} we develop the 
basic theory for the interactions necessary to calculate the capture 
reaction~\nLi. We present the Lagrangian for elastic scattering in the 
$s$- and $p$-wave in the $n+^7$Li system and derive the one-body 
(magnetic moment) and two-body currents. We describe how the couplings 
in the EFT Lagrangian are constrained from available data on low-lying 
bound and resonance states. Results of our calculations are shown 
in Sec.~\ref{sec_crosssection} and compared with some available 
data and potential model calculations. 
We also present a brief discussion on higher-order terms and other degrees 
of freedom that becomes important 
at energies slightly above the ones considered here. 
Finally, we present our conclusions in Sec.~\ref{sec_conclusions}. 
A more technical discussion about the power counting of two-body currents 
in the present approach is given in Appendix~\ref{sec_powercounting}.

\section{Formalism}\label{sec_theory}

The low energy theory for \nLi~ is constructed out of the spin-parity 
$\frac{1}{2}^+$ neutron and $\frac{3}{2}^-$ lithium-7 core. 
The final $^8{\rm Li}$ nucleus contains the $2^+$ and $1^+$ ground and 
excited states in the bound spectrum. 
Very close to the $n$-$^7{\rm Li}$ threshold only the initial 
$s$-wave states are relevant. 
As the energy increases around 200 keV a pronounced $p$-wave resonance 
in the initial state, identified as a $3^+$ state, contributes. 
It is useful then to first list the possible initial and final channels for 
the reaction. 
Concentrating on just the $s$- and $p$-waves, we have in the spectroscopic 
notation $^{2S+1}L_J$: the initial $s$-wave states $^3S_1$ and $^5S_2$, 
the final $p$-wave channels $^3P_2$, $^5P_2$ for the ground state and 
$^3P_1$, $^5P_1$ channels for the 
excited state, and initial $p$-wave resonant state $^5P_3$. 
The $2^+$ ground state of lithium-8 has the quantum numbers of both 
$^3P_2$ and $^5P_2$ states. It is, however, identified with the 
symmetric combination 
$|2^+\rangle\equiv (|^3P_2\rangle+|^5P_2\rangle)/\sqrt{2}$~\cite{Trache:2003}. 
The $1^+$ excited state is primarily dominated by 
the antisymmetric combination 
$|1^+\rangle \equiv (|^5P_1\rangle-
|^3P_1\rangle)/\sqrt{2}$~\cite{Shul'gina1996197}. 
The $3^+$ resonance can only belong to the $^5P_3$ channel in the present 
$n+^7{\rm Li}$ approach. 
The leading contribution to the capture reaction, which comes from the 
initial $^3S_1$ and $^5S_2$ states to the $2^+$ ground state, proceeds 
through E1 transition due to 
electromagnetic selection rules. 
There is a small 
contribution from E1 capture to the excited $1^+$ state 
with a branching ratio of $0.106$~\cite{Lynn:1991}. 
Finally, around 200 keV there is the M1 contribution from the $3^+$ 
resonance to the $2^+$ ground state. 

In Ref.~\cite{Rupak:2011nk}, the authors used the Clebsch-Gordan 
coefficient matrices Eq.~(\ref{eq:ncorespinmatrices})
to project the neutron $N(x)$, lithium-7 $C(x)$ fields onto the spin 
$S=1$ and spin $S=2$ channels as $N^T F_i C$ and $N^T Q_{i j} C$, 
respectively. We use the same formalism here to construct all the 
relevant initial and final states. 

\subsection{Interaction}\label{sec_interaction}
The operators required for the calculation of the capture reaction 
\nLi~ can be classified into three categories: $(a)$ $s$-wave initial 
state elastic scattering $\mathcal L_s$, 
$(b)$ $p$-wave elastic scattering to describe the $3^+$ resonance, 
and analytically continued to describe the $2^+$ ground and 
$1^+$ excited states $\mathcal L_p$, 
and $(c)$ two-body currents that are not related to the elastic channels 
$\mathcal L_\mathrm{EM}= O^{M} +O^{L}$. 

The $s$-wave interaction Lagrangian is written as~\cite{Rupak:2011nk}
\begin{align}\label{eq:Ls}
\mathcal L_s=g^{(1)}(N^T F_i
C)^\dagger(N^T F_i C)+g^{(2)} (N^T Q_{i j} C)^\dagger (N^T Q_{i j} C)+\cdots,
\end{align} 
where the ``$\cdots$" represent higher derivative terms that are suppressed 
at low energy. At LO there is a single coupling 
$g^{(s)}$ in 
$^5S_2$ ($^3S_1$) spin channel that is fixed from the known $s$-wave 
scattering length $a^{(2)}_0= -3.63\pm0.05$ fm 
($a^{(1)}_0=0.87\pm0.07$ fm)~\cite{Koester:1983,*Angulo:2003}. 

As discussed in the next subsection, the description of a low-energy $p$-wave 
bound/excited state or resonance requires two operators at 
LO~\cite{Bertulani:2002sz,Bedaque:2003wa}. 
It is convenient to work in the dimer formalism, where the four-fermion 
contact interaction is replaced by the exchange of an auxiliary dimer 
field $\phi(x)$. The $p$-wave interaction Lagrangian then reads 
\begin{align}\label{eq:Lp}
\mathcal L^{(\eta)}_p={\phi^{(\eta)}_{[j]}}^\dagger 
\left[ \left(i\partial_0+\frac{\nabla^2}{2 M}\right)
+\Delta^{(\eta)}\right]\phi^{(\eta)}_{[j]}+h^{(\eta)}
\left[\phi_{[j]}^{(\eta)} N^T P^{(\eta)} C+\operatorname{H.c.}\right],
\end{align}
where $M=M_n+M_c$, neutron mass $M_n=939.6$ MeV, $^7$Li mass $M_c=6535.4$ MeV 
and H.c. stands for Hermitian conjugate. 
$P^{(\eta)}_{[j]}$ are the projectors for the relevant 
$p$-wave channels: $^3P_1$, $^3P_2$, $^5P_1$, $^5P_2$, $^5P_3$ indicated 
by the index $\eta$ and explicitly given in Appendix~\ref{sec_normalization}. 
The subscript $[j]$ is a single, double or 
triple tensor indices as appropriate for $J=1$, $J=2$ and $J=3$ state, 
respectively. 
The two EFT couplings $\Delta^{(\eta)}$, $h^{(\eta)}$ are proportional to 
the two required operators at LO and are determined from 
elastic scattering data as we discuss later. 

For the capture through the $3^+$ resonance state, we need the magnetic 
moment couplings 
\begin{align}
O^{M}=g_n\mu_N N^T\left(\frac{\bm{\sigma}}{2}\cdot \bm{B}\right)N 
+ g_c\mu_N C^T \left(\bm{J}\cdot \bm{B}\right) C,
\end{align}
where $\bm{\sigma}$ are the Pauli matrices, $\bm{J}$ are the angular 
momentum matrices for spin-3/2 particle, $\bm{B}=\bm{\nabla}\times\bm{A}$ 
the magnetic field, $\mu_N$ the nuclear magneton, and $g_n\equiv 2\kappa_n$ and $g_c=2 \kappa_c/3$ are 
the neutron and the $^7$Li gyromagnetic ratios, respectively. We take 
$\kappa_n=-1.91304$ and $\kappa_c=3.256427$ as the corresponding magnetic 
moments~\cite{Stone:2005}. 
In addition, there are contributions from two-body currents 
\begin{align}\label{eq:LM}
O^{L}=i\mu_N\,L^{(2)}{\phi_{ij}^{(^5P_2)}}^\dagger B_k
\phi^{(^5P_3)}_{lmq}G_{ijklmq}
+\mu_N\,L^{(1)}{\phi_{ij}^{(^3P_2)}}^\dagger B_k\phi^{(^5P_3)}_{lmq}G_{ijklmq}
+{\rm H.c.},
\end{align}
that contribute to M1 transition from initial $^5P_3$ state to final 
$^5P_2$ and $^3P_2$ states, respectively. 
The tensor $G_{ijklmq}$ is defined in Appendix~\ref{sec_normalization}. 
The two-body currents are 
allowed by symmetry so they contribute to the M1 capture. Further, the 
two-body couplings $L^{(1)}$, $L^{(2)}$ 
also regulate the divergences that appear in certain loop diagrams. 
The power counting for the different interactions will be presented in 
details in the following and in Appendix~\ref{sec_powercounting}. 

\subsection{EFT couplings}\label{sec_couplings}
The couplings in Eqs.~(\ref{eq:Ls}), (\ref{eq:Lp}) can be related to 
elastic scattering data in the $s$- and $p$-waves when available. Therefore, 
it is appropriate to match the field theory to the low-energy 
amplitude written in terms of the effective range expansion (ERE). 
In principle, one could incorporate relativistic 
corrections in the EFT amplitude and go beyond the ERE~\cite{Chen:1999tn}. 
However, this is not required at the low energies that we are interested 
in here. The two-body current couplings $L^{(\eta)}$ are not related to the 
elastic scattering data, and are thus determined from the capture data (see 
Sec.~\ref{sec_crosssection}). 

The ERE elastic scattering amplitude in the $\ell$-th partial 
wave is written as 
\begin{align}
i \mathcal A_{\ell}(p)=\frac{2\pi}{\mu} 
\frac{i p^{2 {\ell}}}{p^{2\ell+1} \cot\delta_{\ell}
-ip^{2\ell+1}} 
=\frac{2\pi}{\mu} \frac{i p^{2 {\ell}}}{-1/a_{\ell} + \frac{1}{2}r_{\ell} p^2
+\frac{1}{2} t_{\ell} p^4+ \cdots -ip^{2\ell+1}}, 
\end{align}
with reduced mass $\mu$, and $a_{\ell}$, $r_{\ell}$, $t_{\ell}$, etc., 
ERE parameters. 
Each term is assigned a momentum scaling, in general, given by 
dimensional analysis. 
Since the ERE parameters are associated with the short range (high momentum 
$\Lambda$) nuclear interaction, 
naively one would expect $a_{\ell}\sim\Lambda^{2\ell+1}$, 
$r_{\ell}\sim\Lambda^{2\ell-1}$, $t_{\ell}\sim \Lambda^{2\ell-3}$, and so on. 
At arbitrarily low momentum $p\sim Q$ one can expand the amplitude 
$\mathcal A_{\ell}$ in a Taylor series around $Q/\Lambda=0$, a situation where 
the interaction is weak and perturbative. The more interesting situation 
arises when there are shallow bound or virtual states that one wishes to 
incorporate in the formalism. That implies a rearrangement of the perturbative 
series, which is only possible if at least one of the ERE parameters 
(usually the scattering length) has a different scaling than the one 
assumed by naive dimensional analysis. 
The $s$- and $p$-wave scattering amplitudes at LO then 
read~\cite{Kaplan:1998tg,*Birse:1998dk,*Gegelia:1998gn,*vanKolck:1997ut,
*vanKolck:1998bw,Chen:1999tn,Bertulani:2002sz,Bedaque:2003wa}
\begin{align}\label{eq:ERE_As}
&i\mathcal A_0(p)= \frac{2\pi}{\mu} \frac{i }{-1/a_0 -ip}\ ,\\
&i\mathcal A_1(p) =\frac{2\pi}{\mu} \frac{i p^{2}}{-1/a_1
+\frac{1}{2}r_1 p^2 -ip^{3}}\ .\label{eq:ERE_Ap}
\end{align}
For a shallow bound or virtual state in $s$-wave we set 
$1/a_0\sim Q$, then only one single operator is needed at LO. 
For a shallow $p$-wave the situation is more subtle. First, only one 
particular fine-tuning of the scattering ``length", $1/a_1\sim Q^2\Lambda$, 
is enough to produce the shallow state~\cite{Bedaque:2003wa}. 
However, not only one but two operators emerge at LO, since the effective 
``range" term $r_1p^2/2\sim Q^2\Lambda$ now scales equally as $1/a_1$ at 
momenta 
$p\sim Q$. Second, the unitarity term $ip^3\sim Q^3$ is in principle of 
higher order. The $p$-wave amplitude ${\cal A}_1$ is then suppressed by 
$Q/\Lambda$ relative to ${\cal A}_0$. 
However, for energies close to the resonant state there is a cancellation 
of the leading terms (kinematical 
fine-tuning~\cite{Bedaque:2003wa,Pascalutsa:2002pi}) that 
makes $-1/a_1+rp^2/2\sim Q^3$ and promotes the unitarity term 
$ip^3$ to LO. In that region, the amplitudes (\ref{eq:ERE_As}) and 
(\ref{eq:ERE_Ap}) contribute at the same order. 

\begin{figure}[tbh]
\begin{center}
\includegraphics[width=0.47\textwidth,clip=true]{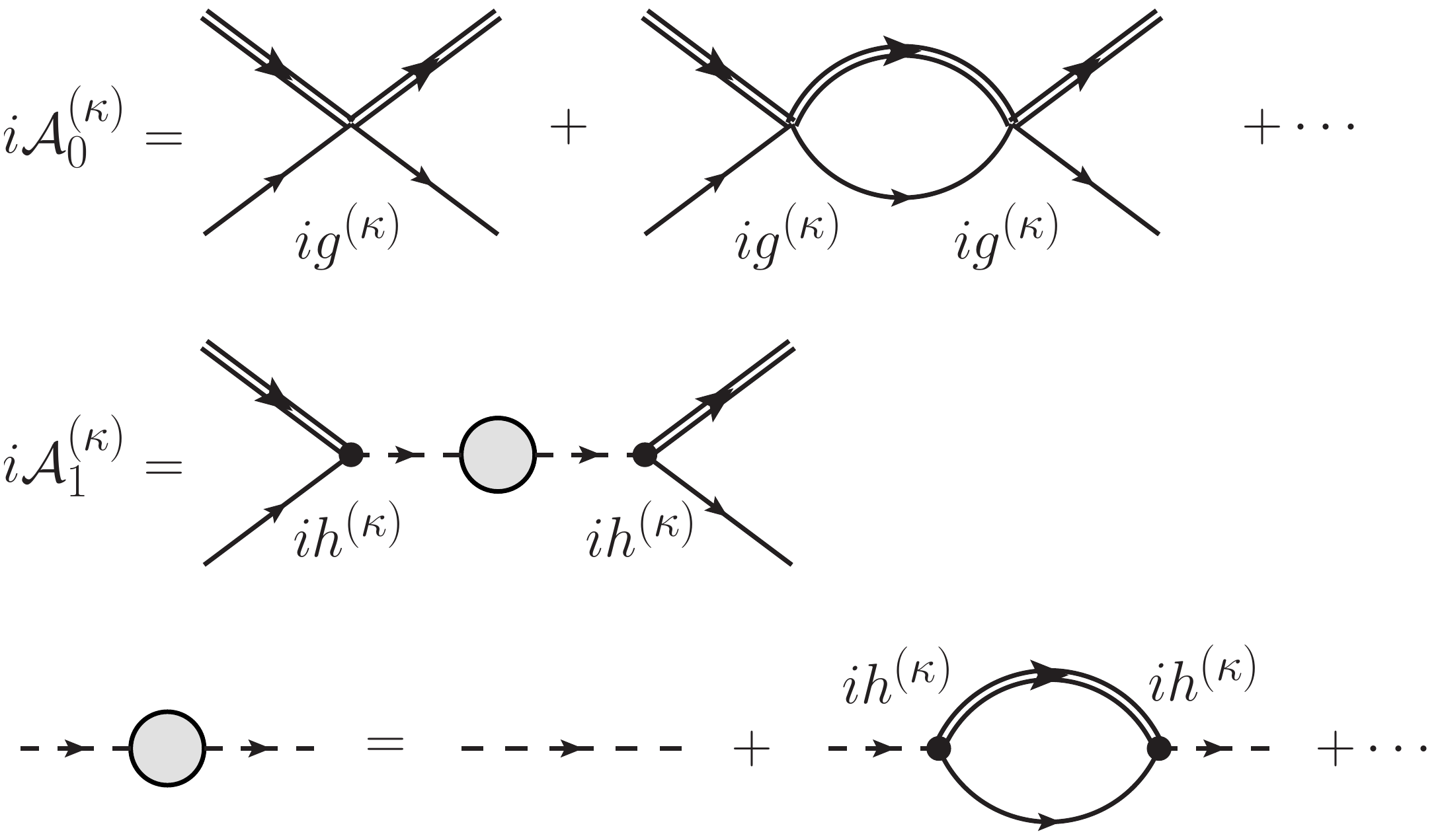} 
\end{center}
\caption{\protect Initial $s$-wave $\mathcal A_0^{(\kappa)}$ 
and $p$-wave $\mathcal A_1^{(\kappa)}$ elastic scattering amplitudes. 
Double line is the $^7$Li propagator, single line the neutron propagator, 
dashed line the bare dimer propagator.}
\label{fig:scattering}
\end{figure}

The elastic scattering amplitude in EFT is calculated from the interactions 
in Eqs.~(\ref{eq:Ls}), (\ref{eq:Lp}) as shown in Fig.~\ref{fig:scattering}. 
We get for the $s$-wave amplitude 
\begin{align}\label{eq:EFT_As}
&i\mathcal A^{(\eta)}_0(p)=\frac{i g^{(\eta)}}{1-i g^{(\eta)} f_0(p)},
\\
&f_0(p)= -i 2\mu\(\frac{\lambda}{2}\)^{4-D}\int 
\frac{d^{D-1}\bm{q}}{(2\pi)^{D-1}}\frac{1}{q^2- p^2-i 0^+}= 
-\frac{i\mu}{2\pi}(\lambda+ip), \nonumber 
\end{align} 
with renormalization scale $\lambda$. Comparing Eqs.~(\ref{eq:EFT_As}) and 
(\ref{eq:ERE_As}) we find~\cite{Kaplan:1998tg,*Birse:1998dk,*Gegelia:1998gn,
*vanKolck:1997ut,*vanKolck:1998bw,Chen:1999tn} 
\begin{align}
g^{(\eta)}(\lambda)= \frac{2\pi}{\mu}\frac{1}{\lambda-1/a_0^{(\eta)}},
\end{align}
where $\eta=1,2$ correspond to spin channels $^3S_1$, $^5S_2$, respectively. 

For the $p$-wave amplitude we have from Fig.~\ref{fig:scattering} 
\begin{align}\label{eq:EFT_Ap}
&i\mathcal A^{(\eta)}_1=-[ h^{(\eta)}]^2 \frac{ p^2}{\mu^2} 
iD^{(\eta)}(p^2/2\mu,0) = \frac{2\pi}{\mu} 
\frac{i p^2}{-\frac{2\pi\mu\Delta^{(\eta)}}{[h^{(\eta)}]^2}
-\frac{\pi\lambda^3}{2} -\left(\frac{3 \lambda}{2} 
+\frac{\pi}{[h^{(\eta)}]^2}\right)p^2 -i p^3} ,
\end{align}
using the full dimer propagator 
\begin{align}\label{eq:dimer}
i D^{(\kappa)}(p_0,\bm{p})=& \frac{i }
{\Delta^{(\kappa)} - \frac{1}{2\mu}\zeta^2 +\frac{2h^{(\kappa)\,2}}{\mu}
f_1 (p_0,\bm{p})},\\
f_1(p_0,\bm{p})=&
\frac{1}{4\pi}\left(\zeta^3-\frac{3}{2}\zeta^2\lambda
+\frac{\pi}{2} \lambda^3 \right), \nonumber
\end{align}
where $\zeta=\sqrt{-2\mu p_0 +\mu p^2/M -i 0^+}$. 
Matching the EFT amplitude to the $p$-wave ERE expansion 
determines the coupling pair ($\Delta^{(\kappa)}$, $h^{(\kappa)}$). 
As in the $s$-wave, comparing Eqs.~(\ref{eq:EFT_Ap}) and (\ref{eq:ERE_Ap}) 
yields 
\begin{align}
-\frac{2\pi\mu\Delta^{(\eta)}}{[h^{(\eta)}]^2}-\frac{\pi}{2}\lambda^3=&
-1/a_1^{(\eta)}, 
\nonumber\\
-\frac{3}{2}\lambda-\frac{\pi}{[h^{(\eta)}]^2}=&\frac{1}{2} r_1^{(\eta)}.
\label{eq:a1r1}
\end{align}
The index $\eta$ identifies the relevant $p$-wave channels outlined earlier. 
The EFT couplings $\Delta^{(\eta)}$, $h^{(\eta)}$ are therefore fixed 
in terms of the ERE parameters $a_1^{(\eta)}$, $r_1^{(\eta)}$ 
and the renormalization scale $\lambda$. 

In the $^5P_3$ resonance channel we call $\eta=3$ and directly use the 
relations in Eq.~(\ref{eq:a1r1}). The parameters $a_1^{(3)}$ and $r_1^{(3)}$ 
are related to the known resonance position and width, as shown in 
Sec.~\ref{sec_resonance}. 

For the bound channels $1^+$, $2^+$, we follow the procedure used in 
Ref.~\cite{Rupak:2011nk}. It is more convenient to work directly with the 
location of the pole in the dimer propagator at the binding momentum 
$\gamma^{(\eta)}$ and its residue ${\cal Z}^{(\eta)}$. 
The latter is the wave function renormalization constant calculated using 
Eqs.~(\ref{eq:dimer}), (\ref{eq:a1r1}) as
\begin{align}
[{\cal Z}^{(\eta)}]^{-1} =& \frac{\partial}{\partial p_0} 
[D^{(\eta)}(p_0,\bm{p})]^{-1}\big|_{p_0= \bm{p}^2/(2M) -B^{(\eta)}},
\nonumber\\
{\cal Z}^{(\eta)}=& -\frac{2\pi}{[h^{(\eta)}]^2}\frac{1}{3\gamma^{(\eta)}
+r_1^{(\eta)}},
\end{align} 
where $B^{(\eta)}=[\gamma^{(\eta)}]^2/(2 \mu)$ is the binding energy. 
The $^3P_2$ and $^5P_2$ channels in $2^+$ share a common binding momentum 
$\gamma^{(2^+)}\approx 57.8$ MeV. Moreover, the capture cross section is not 
independently sensitive to the effective range parameter $r_1$ in these two 
spin channels~\cite{Rupak:2011nk}. 
For this reason we use a common effective range 
parameter $r_1^{(2^+)}$. This coincidentally gives the 
observed 
ratio $0.82$ of spin channel $S=2$ relative to the total E1 capture to the 
ground state at threshold~\cite{Barker:1995a}. 
We make a similar simplifying assumption 
for the $1^+$ state and use a common effective range parameter $r_1^{(1^+)}$ 
for both spin channels $^3P_1$, $^5P_1$. In the final cross section only the 
combinations 
\begin{align}
[h^{(2^+)}]^2 {\cal Z}^{(2^+)} &=
-\frac{2\pi}{3\gamma^{(2^+)}+r_1^{(2^+)}},\nonumber\\
[h^{(1^+)}]^2 {\cal Z}^{(1^+)} &=
-\frac{2\pi}{3\gamma^{(1^+)}+r_1^{(1^+)}},
\end{align}
contribute to the $2^+$ and $1^+$ states, respectively. In 
Ref.~\cite{Rupak:2011nk}, $r_1^{(2^+)}\approx -1.55$ fm$^{-1}$ from a fit to 
low energy data from Ref.~\cite{Blackmon:1996}. 
$\gamma^{(1^+)}\approx 41.6$ MeV from the known $1^+$ excited state energy. 
We will determine $r_1^{(2^+)}$ and $r_1^{(1^+)}$ in this work from the known 
E1 thermal capture rates to the $2^+$ and $1^+$ states~\cite{Lynn:1991}, 
respectively. 

\subsection{Resonance parameters}\label{sec_resonance}

The $p$-wave EFT couplings for the $3^+$ ($^5P_3$) state can be related 
to the ERE scattering parameters $a_1^{(3)}$ and $r_1^{(3)}$ through 
Eq~(\ref{eq:a1r1}). Since these 
ERE parameters are not known we first determine them from the known 
resonance energy and width. At very low momentum,
the phase shift $\delta(p)$ in this channel vanishes 
due to the centrifugal barrier. 
Near the resonance, $\delta(p)$ increases rapidly through $\pi/2$ 
from below. Thus, $\cot\delta (p)$ has 
to go through zero from above, {\em i.e.}, it has a negative slope. 
We write these conditions as~\cite{Sakurai} 
\begin{align}\label{eq:ErGamma}
\cot\delta\Big|_{E=E_r} =0,\ \ \mathrm{and}\\
\frac{d}{d E}\cot\delta\Big|_{E=E_r} = - c <0, \nonumber
\end{align}
such that $c>0$. Near the resonance
\begin{align}
\cot\delta (E)\approx &\cot\delta(E_r) +(E-E_r)\cot'\delta(E_r) 
= -(E-E_r) c,\\
\mathcal A(p)=&\frac{2\pi}{\mu}\frac{1}{p\cot\delta -i p} \approx 
\frac{2\pi}{\mu p}\frac{-{\Gamma_r}/{2}}{(E-E_r)+i{\Gamma_r}/{2}}, \nonumber
\end{align}
and we recover the Breit-Wigner form, identifying $c\equiv 2/\Gamma_r$. 
The resonance position and width can be related to the ERE scattering 
parameters. 
In the center-of-mass (c.m.) frame we define $p_r^2 \equiv 2\mu E_r$, then 
\begin{align}
p^3\cot\delta = -\frac{1}{a_1^{(3)}}+\frac{r_1^{(3)} p^2}{2}\Rightarrow 
\cot\delta(p_r) = -\frac{1}{a_1^{(3)} p_r^3} +\frac{r_1^{(3)}}{2 p_r} =0,
\ \ \mathrm{and}\\ 
\frac{d}{d E}\cot\delta(E_r) = \frac{\mu}{p_r} \frac{d}{d p_r}\cot\delta(p_r)=
-2/\Gamma_r \Rightarrow \frac{3}{a_1^{(3)} p_r^4} -\frac{r_1^{(3)}}{2p_r^2}=
-\frac{2 p_r}{\mu\Gamma_r}\nonumber.
\end{align}
For the resonance momentum and width one gets 
(see also Refs.~\cite{Higa:2008dn,*Gelman:2009be,Long:2009wq}) 
\begin{equation}
p_r^2=\frac{2}{a_1^{(3)}r_1^{(3)}}\,,\qquad\mbox{and}\qquad
\Gamma_r=-\frac{2p_r^3}{\mu r_1^{(3)}}\,.
\end{equation}
Solving these equations we find
\begin{align}\label{eq:ResonanceParameters}
a_1^{(3)}=-\frac{\mu\Gamma_r}{p_r^5},\ \ \mathrm{and}\ \ r_1^{(3)}=
-\frac{2p_r^3}{\mu\Gamma_r}. 
\end{align}
We choose $p_r=+\sqrt{2\mu E_r}$ so that $(a_1^{(3)} ,r_1^{(3)})$ are 
negative to be consistent with the Wigner 
bound~\cite{Hammer:2009zh,*Hammer:2010fw}. 
If we take the experimental central values for $E_r=0.222$ MeV and 
$\Gamma_r=0.031$ MeV in the c.m. frame 
we find $(|1/a_1^{(3)}|^{1/3}, r_1^{(3)})=(46.38, -547.1 )$ MeV, and from 
Eq~(\ref{eq:a1r1}) we are able to determine the EFT couplings. 

\begin{figure}[tbh]
\begin{center}
\includegraphics[width=0.48\textwidth,clip=true]{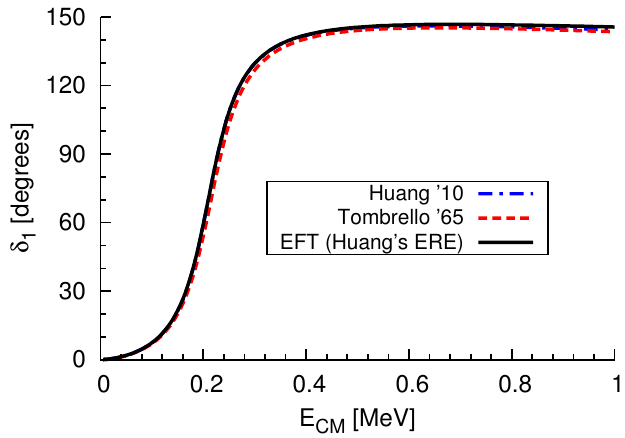} 
\includegraphics[width=0.48\textwidth,clip=true]{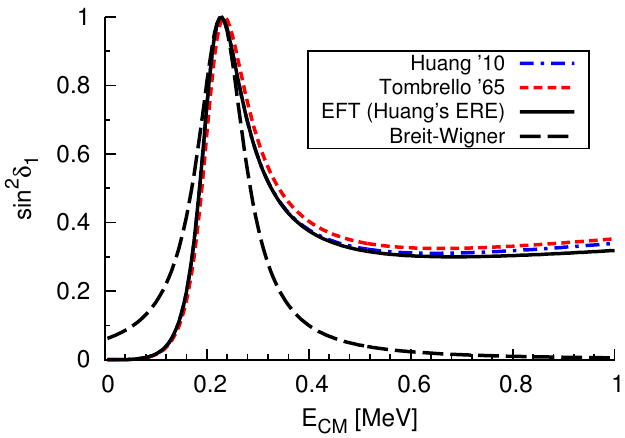} 
\end{center}
\caption{\protect $^5P_3$ phase shifts. 
Blue dot-dashed curve using Huang's~\cite{Huang:2008ye} Woods-Saxon 
parameters, red dashed curve using Tombrello's~\cite{Tombrello:1965} 
parameters, black solid curve EFT fitted to Huang's 
``synthetic data'', black long-dashed curve Breit-Wigner form. }
\label{fig:Resonance}
\end{figure}

To test the robustness of the above procedure we generate ``synthetic data'' 
for the resonance phase shifts from a known nuclear interaction. 
We take a single-particle potential given by the Woods-Saxon 
form~\cite{Huang:2008ye}, 
\begin{align}
V(r)= -v_0 \left[1+\exp\left( \frac{r-R_c}{a_c}\right)\right]^{-1}
+\frac{1}{2}\frac{v_{so}}{r}\frac{d}{dr}
\left[1+\exp\left( \frac{r-R_c}{a_c}\right)\right]^{-1},
\end{align}
where the $1/2$ factor in the second term comes from the 
expectation value of the single-particle spin-orbit operator 
in the $3^+$ channel. 
We use natural units with $\hbar=1=c$. 
In a study by Tombrello~\cite{Tombrello:1965}, the central 
potential with a depth $v_0=26.42$ MeV, spin-orbit $v_{so}=0$, 
range $R_c=2.95$ fm and diffusiveness 
$a_c=0.52$ fm was used to reproduce the resonance energy. 
A more recent work from Huang et al.~\cite{Huang:2008ye} uses 
$v_0=34.93$ MeV, spin-orbit potential depth $v_{so}=10$ MeV, 
$R_c=2.391$ fm, and $a_c=0.65$ fm. 
The two sets of parameters produce nearly identical 
phase shifts, Fig.\ref{fig:Resonance}. 
Plugging Huang's synthetic data into Eq.~(\ref{eq:ErGamma}) generates 
$E_r=0.228$ MeV and $\Gamma_r=0.115$ MeV. 
From Eq.~(\ref{eq:ResonanceParameters}) one gets 
$(|1/a_1^{(3)}|^{1/3}, r_1^{(3)})=(30.69, -154.3)$ MeV, which are then 
used as input to the EFT curve shown in Fig.\ref{fig:Resonance}. 
Alternatively, one can extract directly the ERE parameters from the 
low-energy behavior of the phase shifts: 
$(|1/a_1^{(3)}|^{1/3}, r_1^{(3)})=(31.02, -157.6 )$ MeV and 
$(|1/a_1^{(3)}|^{1/3}, r_1^{(3)})=(30.84, -158.9 )$ MeV in the case of 
Tombrello and Huang's parameters, respectively. We also show the 
Breit-Wigner curve using the extracted resonance parameters. 
We see that the procedure 
outlined allows EFT to reproduce well the synthetic data at low energy. 

We further note that the potential model parameters chosen give a 
resonance width that is about three times 
larger than the experimental value. 
As we discuss in Sec.~\ref{sec_crosssection}, it does impact the shape of 
the M1 curve in the capture reaction. 
To fit the resonance width to experimental 
data would require tuning yet another parameter (besides the potential depth) 
in the Woods-Saxon potential. 
In the EFT language, two operators are required to describe a shallow 
$p$-wave resonance at LO. The corresponding couplings 
$(\Delta^{(3)},h^{(3)})$ are directly related to two experimental 
data: $E_r$ and $\Gamma_r$.

\section{Capture Cross section}\label{sec_crosssection}

The $^7{\rm Li}(n,\gamma)^8{\rm Li}$ capture from $E=0$ to about 
$0.15$ MeV total c.m. energy is almost entirely given by the E1 transition 
from initial $s$-waves. The dominant contribution goes directly to the 
$2^+$ ground state, with a small fraction (about 10\%) going to the 
$1^+$ excited state. 
In a recent work~\cite{Rupak:2011nk} the former was calculated explicitly, 
but the later was only taken into account implicitly with the use of the 
experimental branching ratio. 
Around $E\approx 200$ keV the $3^+$ resonance 
enhances the cross section through an M1 transition to the ground state. 
In this section we derive expressions for these three low-energy 
mechanisms and compare our results with potential model calculations 
and few available experimental data. 

The \nLi~ cross section is calculated in the c.m. frame 
with $\bm{p}$ ($\bm{k}$) the core (photon) momentum and 
$\hat{\bm{k}}\cdot\hat{\bm{p}}=\cos\theta$. The incoming momentum $p$ 
as well as the binding momenta for the ground and excited states 
$\gamma^{(\eta)}$ are assumed to be of the order of the low-energy scale, 
{\em i.e.}, $p\sim\gamma^{(\eta)}\sim Q$. The photon at LO has 
$|\bm{k}|=k_0\approx (p^2+[\gamma^{(\eta)}]^2)/(2\mu)$ and the Mandelstam 
variable $s\approx (M_n+M_c)^2=M^2$. 
The c.m. differential cross section is 
\begin{align}
\frac{d\sigma}{d\phi d \cos\theta}=\frac{1}{64\pi^2 s}
\frac{|\bm{k}|}{|\bm{p}|}|\mathcal M|^2
\approx \frac{1}{64\pi^2 M^2}\frac{p^2+[\gamma^{(\eta)}]^2}{2\mu p}|
\mathcal M|^2,
\end{align}
where $\gamma^{(2^+)}\approx 57.8$ MeV and 
$\gamma^{(1^+)}\approx 41.6$ MeV for the $2^+$ and $1^+$ final 
states, respectively. 

\begin{figure}[tbh]
\begin{center}
\includegraphics[width=0.47\textwidth,clip=true]{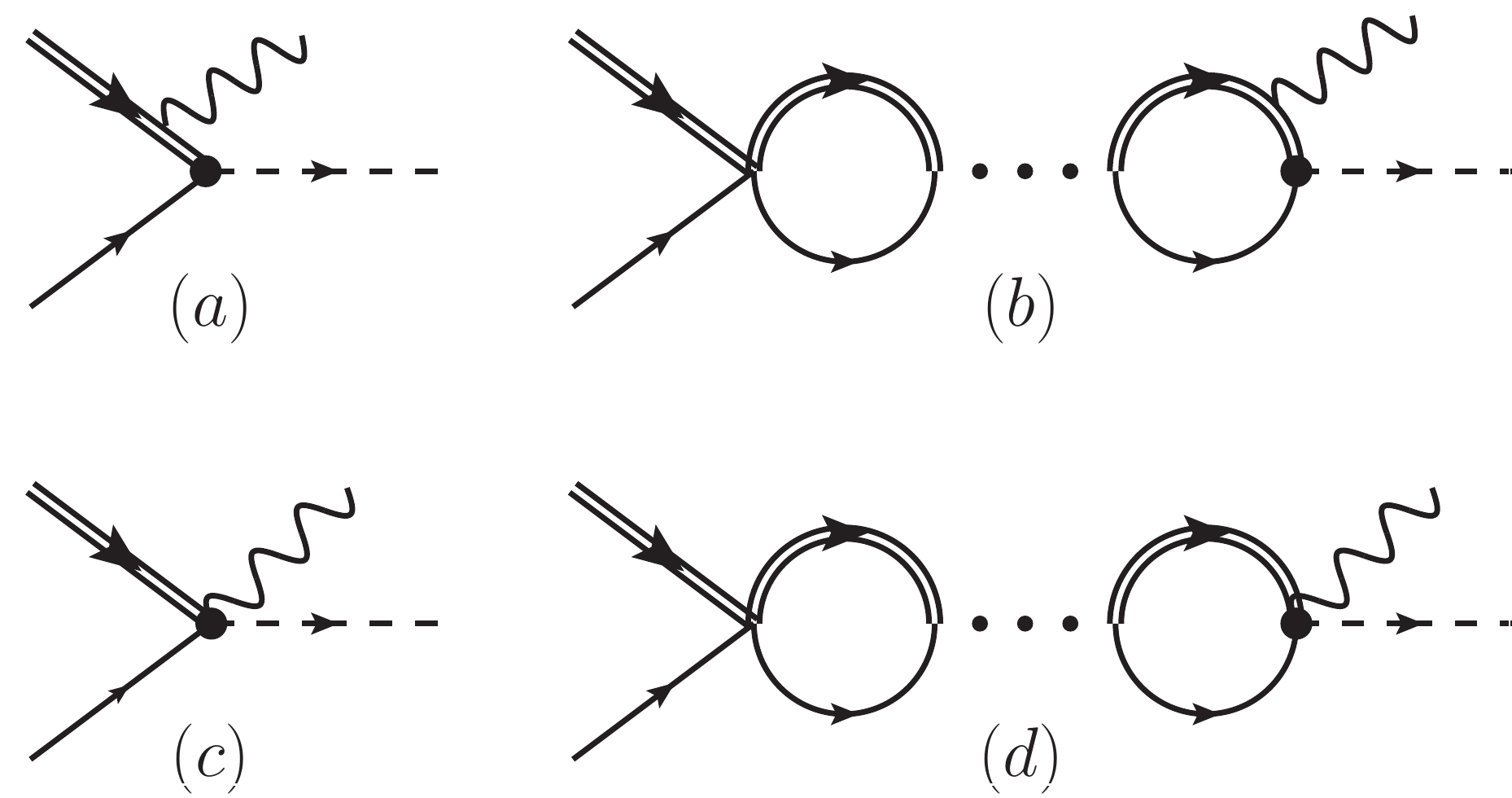} 
\end{center}
\caption{\protect Capture reactions \nLi. Wavy lines represent photons. 
$``\cdots"$ represents initial state $s$-wave interaction. }
\label{fig:E1capture}
\end{figure}

The E1 capture to the ground state at LO proceeds through 
the diagrams in Fig.~\ref{fig:E1capture}. 
The photons are minimally coupled to the $^7$Li nucleus by gauging the 
core derivatives $\bm{\nabla}_C\rightarrow \bm{\nabla}_C+ieZ_c\bm{A}$ 
with $Z_c=3$ the core charge. We quote the final result for capture to 
the $^5P_2$ state from 
Ref.~\cite{Rupak:2011nk} here for completeness: 
\begin{eqnarray}
\big|\mathcal M^{(^5P_2)}_{\rm E1}\big|^2&=&
5 \left(\frac{Z_c M_n}{M}\right)^2
\frac{64\pi \alpha M^2\Big|h^{(2^+)} \sqrt{{\cal Z}^{(2^+)}}\Big|^2}{ \mu}
\nonumber \\&&
\times\[ |1\!+\!X|^2\!-\!\frac{p^2\sin^2\theta}{p^2 +[\gamma^{(2^+)}]^2}
\(\frac{2[\gamma^{(2^+)}]^2}{p^2+[\gamma^{(2^+)}]^2}\!+\!X\!+\!X^\ast\) \], 
\label{eq:5p2capture}
\\[3.0mm]
X&=&\frac{1}{-1/a^{(2)}_0-i p}\left[\frac{2}{3}
\frac{[\gamma^{(2^+)}]^3-ip^3}{[\gamma^{(2^+)}]^2+p^2}+ip\right],
\nonumber
\end{eqnarray}
with $\alpha=e^2/(4\pi)$. 
The amplitude $\big|\mathcal M^{(^3P_2)}_{\rm E1}\big|$ 
is obtained from the above by replacing $a_0^{(2)}\rightarrow a_0^{(1)}$. 
The E1 transition to the $2^+$ ground state is written as 
\begin{align}\label{eq:totalsigmaE1gs}
\frac{d\sigma^{(2^+)}_{\rm E1}}{d \cos\theta}
=\frac{1}{32\pi M^2}
\frac{p^2+[\gamma^{(2^+)}]^2}{2\mu p}\frac{1}{8}
\frac{\big|\mathcal M^{(^5P_2)}_{\rm E1}\big|^2
+\big|\mathcal M^{(^3P_2)}_{\rm E1}\big|^2}{2}, 
\end{align}
taking the $^8$Li ground state $|2^+\rangle$ as the symmetric combination 
of final states. 
The total cross section $\sigma(p)$ is obtained from a straightforward 
integration over the angle $\theta$. We have used a common effective range 
parameter $r_1^{(2^+)}$ in $\mathcal{Z}^{(2^+)}$ as explained earlier, 
see Ref.~\cite{Rupak:2011nk} for details.  Fitting to the thermal cross 
section $\sigma^{(2^+)}=40.56$ mb~\cite{Lynn:1991} gives 
$r_1^{(2^+)}=-1.47$ fm$^{-1}$, which is close to the 
$r_1^{(2^+)}=-1.55$ fm$^{-1}$ value one gets using the data from 
Ref~\cite{Blackmon:1996}. In this work we will use the value obtained 
from the thermal capture. 

\begin{figure}[tbh]
\begin{center}
\includegraphics[width=0.55\textwidth,clip=true]{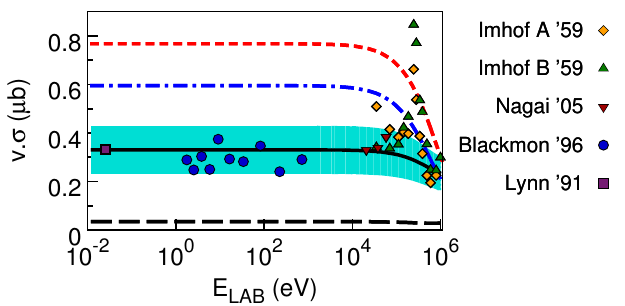} 
\end{center}
\caption{\protect 
Black long-dashed and solid curves are the EFT results for the 
E1 capture to the excited state and the total E1 capture, respectively. 
The shaded area shows the estimated $30\%$ EFT errors in the latter. 
The results of the potential model code CDXS+~\cite{CDXSp} using parameters 
from Ref.~\cite{PhysRevC.68.045802} and Ref.~\cite{Tombrello:1965} are 
given respectively by the blue dot-dashed and red dashed curves. The experimental 
points are from Refs.~\cite{Imhof:1959,Nagai:2005,Blackmon:1996,Lynn:1991}.}
\label{fig:E1captureB}
\end{figure}

The E1 capture cross section to the $1^+$ excited state, 
$\sigma^{(1^+)}_{\rm E1}$, comprises the same set of diagrams in 
Fig.~\ref{fig:E1capture} except the final state dimer, which is in the 
$1^+$ state. The sum over the final polarization states gives an 
overall factor of $3$ instead of $5$. Therefore we have 
\begin{eqnarray}
\big|\mathcal M^{(^5P_1)}_{\rm E1}\big|^2&=&
3 \left(\frac{Z_c M_n}{M}\right)^2
\frac{64\pi \alpha M^2\Big|h^{(1^+)} \sqrt{{\cal Z}^{(1^+)}}\Big|^2}{ \mu}
\nonumber \\&&
\times\[ |1\!+\!Y|^2\!-\!\frac{p^2\sin^2\theta}{p^2 +[\gamma^{(1^+)}]^2}
\(\frac{2[\gamma^{(1^+)}]^2}{p^2+[\gamma^{(1^+)}]^2}\!+\!Y\!+\!Y^\ast\) \], 
\label{eq:5p1capture}
\\[3.0mm]
Y&=&\frac{1}{-1/a^{(2)}_0-i p}\left[\frac{2}{3}
\frac{[\gamma^{(1^+)}]^3-ip^3}{p^2+[\gamma^{(1^+)}]^2}+ip\right],
\nonumber\\[3.0mm]
\frac{d\sigma^{(1^+)}_{\rm E1}}{d \cos\theta}
&=&\frac{1}{32\pi M^2}
\frac{p^2+[\gamma^{(1^+)}]^2}{2\mu p}\frac{1}{8}
\frac{\big|\mathcal M^{(^5P_1)}_{\rm E1}\big|^2
+\big|\mathcal M^{(^3P_1)}_{\rm E1}\big|^2}{2},
\label{eq:totalsigmaE1excited}
\end{eqnarray}
where we made the replacements 
$[h^{(2^+)}]^2\mathcal{Z}^{(2^+)}\rightarrow 
[h^{(1^+)}]^2\mathcal{Z}^{(1^+)}$, 
$\gamma^{(2^+)}\rightarrow\gamma^{(1^+)}$ in 
Eqs.~(\ref{eq:5p1capture},\ref{eq:totalsigmaE1excited}). 
The $|1^+\rangle$ state is considered the anti-symmetric combination of the 
final states. 
As in the $2^+$ case, the amplitude in the other channel spin 
$\big|\mathcal M^{(^3P_1)}_{\rm E1}\big|$ is derived from 
$a_0^{(2)}\rightarrow a_0^{(1)}$, with a common effective range parameter 
$r_1^{(1^+)}$.  From the thermal capture rate $\sigma^{(1^+)}=4.80$ mb to 
the $1^+$ state we get $r_1^{(1^+)}\approx -1.93$ fm$^{-1}$. 
The E1 capture cross section to the ground and excited state is shown 
in Fig.~\ref{fig:E1captureB}. 
We also show the potential model results for comparison. 
The leading uncertainty in potential model results seems to be associated 
with the poorly known effective range $r_1^{(2^+)}$ that we determine from 
the thermal capture rate. We also notice that the data set that we call 
ImhofB is more consistent with the $1/v$ behavior suggested by the 
Blackmon~\cite{Blackmon:1996} and Lynn~\cite{Lynn:1991} data than ImhofA. 

\begin{figure}[tbh]
\begin{center}
\includegraphics[width=0.5\textwidth,clip=true]{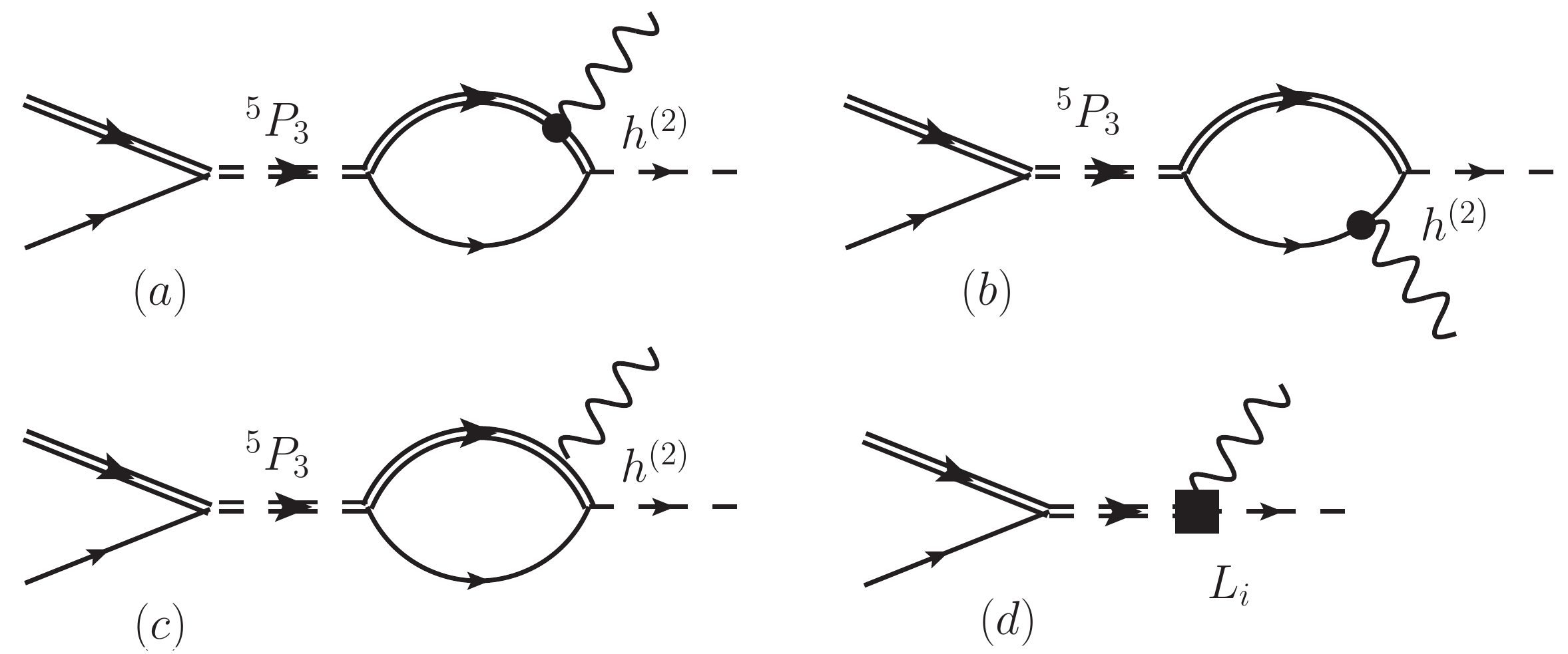} 
\end{center}
\caption{\protect The photon is coupled through the magnetic moment to the 
neutron (single solid line) and $^7$Li nucleus (double solid line). Single 
dashed line is the ground state $^8$Li $2^+$ dressed dimer, double dashed 
line is the $^8$Li $3^+$ resonance dressed dimer. }
\label{fig:3plusCapture}
\end{figure}

Next, we consider the M1 capture cross section. It proceeds through an 
initial $p$-wave state,  and therefore, suppressed at low energies. 
In contrast, the E1 capture takes place via initial $s$-wave states 
and displays the known $1/p$ enhancement at low momentum. 
However, the presence of the $3^+$ resonance enhances the 
M1 contribution around the resonance energy, making it comparable to E1. 
In Fig.~\ref{fig:3plusCapture} we show the diagrams that make the 
leading contributions to the M1 capture. 
The first two involve the neutron and core magnetic moment couplings, 
and contribute to both $^5P_2$ and $^3P_2$ final states. 
In the third one the magnetic photon couples to the charged $^7{\rm Li}$ 
core ``in flight'' or, in a more classical picture, to the 
electromagnetic current generated by the orbital motion of the 
charged $^7{\rm Li}$ core. 
It arises from minimal photon coupling and contributes only to the 
$^5P_2$ final state. 
The last diagram contains two-body currents in the respective $^5P_2$ 
and $^3P_2$ channels. Naively, counting only factors of $Q/\Lambda$, the 
contributions from two-body currents seem to be more important than the ones 
from the magnetic moments. 
We will come back to this point in the following. 

Evaluation of these diagrams in the final $^5P_2$ channel yields 
\begin{align}
i\mathcal M^{(^5P_2)}_{\rm M1}=&
\epsilon_{abc}\varepsilon_{ij}^\ast\varepsilon^{(\gamma)\ast}_c G_{ijalmq} 
p_q k_b\, h^{(2^+)} \sqrt{2M\mathcal{Z}^{(2^+)}}\,\mu_N
\bigg[-\frac{[h^{(3)}]^2}{2\pi\mu} D^{(3)}(p^2/2\mu,0)\bigg]\sqrt{3}
\nonumber\\&\times
U^T_N(-\bm{p}) Q_{l m} U_C(\bm{p})
\bigg[\sqrt{\frac{3}{2}}\bigg(\frac{3}{2}g_c\!+\!\frac{1}{2}g_n
\!+\!\frac{2\mu Z_cM_n}{M_c^2}\bigg) I(p;\lambda)
\!-\!\frac{2\pi L^{(2)}}{h^{(3)} h^{(2^+)}}\bigg],
\label{eq:resonancecaptureA}
\end{align}
with 
\begin{equation}
I(p;\lambda)=\frac{2}{3}
\frac{[\gamma^{(2^+)}]^3-i p^3}{[\gamma^{(2^+)}]^2+p^2}-\lambda.
\label{eq:Iintegral}
\end{equation}
The various tensors above are defined in Appendix~\ref{sec_normalization}. 
For the $^3P_2$ channel one gets 
\begin{align}
i\mathcal M^{(^3P_2)}_{\rm M1}=&
-i\epsilon_{abc}\varepsilon_{ij}^\ast\varepsilon^{(\gamma)\ast}_c G_{ijalmq} 
p_q k_b\, h^{(2^+)} \sqrt{2M\mathcal{Z}^{(2^+)}}\,\mu_N
\bigg[-\frac{[h^{(3)}]^2}{2\pi\mu} D^{(3)}(p^2/2\mu,0)\bigg]\sqrt{3}
\nonumber\\&\times
U^T_N(-\bm{p}) Q_{l m} U_C(\bm{p})
\bigg[\sqrt{\frac{3}{2}}\bigg(\frac{3}{2}g_c\!-\!\frac{3}{2}g_n
\bigg) I(p;\lambda)
\!-\!\frac{2\pi L^{(1)}}{h^{(3)} h^{(2^+)}}\bigg].
\label{eq:resonancecaptureB}
\end{align}
Numerically the gyromagnetic factors are 
$\sqrt{3/2} (3 g_c/2+g_n/2+2\mu Z_c M_n/M_c^2)\sim 1.8$ for $^5P_2$ and 
$\sqrt{3/2} (3 g_c/2-3g_n/2)\sim 11$ for $^3P_2$. 
The former is of natural size for a dimensionless constant ---in 
Eq.~(\ref{eq:resonancecaptureA}), the two-body current dominates for 
a natural $L^{(2)}\sim 1$ and the loop contribution is subleading. 
However, the $^3P_2$ numerical factor is large and enhances the loop 
contribution beyond the estimates of the power counting. 
Thus in Eq.~(\ref{eq:resonancecaptureB}) 
the loop contribution is as important as the two-body current 
and both enter at leading order. 
For convenience, we keep the loop contribution and two-body magnetic 
coupling at the same order in both the spin channels. 
The dependence on the renormalization scale $\lambda$ coming from the loop 
function $I(p,\lambda)$, Eqs.~(\ref{eq:resonancecaptureA}), 
(\ref{eq:resonancecaptureB}), is then cancelled by the two-body 
couplings $L^{(1)}$ and $L^{(2)}$, respectively. We write 
\begin{eqnarray}
L^{(2)}&=&-\frac{h^{(3)} h^{(2)}}{2\pi}\bigg[\sqrt{\frac{3}{2}}\bigg(
\frac{3}{2}g_c+\frac{1}{2}g_n+\frac{2\mu Z_cM_n}{M_c^2}\bigg)\lambda
+\beta^{(2)}\bigg],
\nonumber\\
L^{(1)}&=&-\frac{h^{(3)} h^{(2)}}{2\pi}\bigg[\sqrt{\frac{3}{2}}\bigg(
\frac{3}{2}g_c-\frac{3}{2}g_n\bigg)\lambda+\beta^{(1)}\bigg],
\end{eqnarray}
where $\beta^{(i)}$'s are 
renormalized two-body parameters with dimensions of mass. 
The total cross section can be written as 
\begin{eqnarray}
\sigma^{(2^+)}_{\rm M1}&=&\frac{1}{14}\frac{7}{3}\frac{\alpha \mu}{M_p^2}
[h^{(2^+)}]^2\big|\mathcal{Z}^{(2^+)}\big|
\frac{(p^2+[\gamma^{(2^+)}]^2)^3}{(2\mu p)^3} 
\bigg|\frac{p^2}{-1/a_1^{(3)}+\frac{1}{2}r_1^{(3)} p^2-ip^3}\bigg|^2
\nonumber\\&&\times
\Bigg\{
\bigg|\sqrt{\frac{2}{3}}\bigg(\frac{3}{2}g_c+\frac{1}{2}g_n
+\frac{2\mu Z_cM_n}{M_c^2}\bigg)\frac{[\gamma^{(2^+)}]^3-ip^3}
{[\gamma^{(2^+)}]^2+p^2}+\beta^{(2)}\bigg|^2
\nonumber\\&&
+\bigg|\sqrt{\frac{2}{3}}\bigg(\frac{3}{2}g_c-\frac{3}{2}g_n\bigg)
\frac{[\gamma^{(2^+)}]^3-ip^3}{[\gamma^{(2^+)}]^2+p^2}
+\beta^{(1)}\bigg|^2\Bigg\},
\label{eq:M1sigma}
\end{eqnarray}
where the proton mass $M_p=938.3$ MeV is used. 
We summed over the final state dimer and photon polarizations, and 
averaged over the initial spin states. 
The magnetic moment and orbital 
momentum weights are easy to understand if one compares with the 
non-relativistic quantum operator for the M1 transition, which is 
proportional to 
\begin{align}
\left(\frac{\mu M_n Z_c}{M_c^2} \vec{\bm{L}} +g_c \vec{\bm{S}}_C 
+g_n \vec{\bm{S}}_N\right)_z,
\end{align}
and its expectation value between the initial $^5P_3$ state and the final 
$^5P_2$ and $^3P_2$ states, respectively. 

\begin{figure}[tbh]
\begin{center}
\includegraphics[width=0.48\textwidth,clip=true]{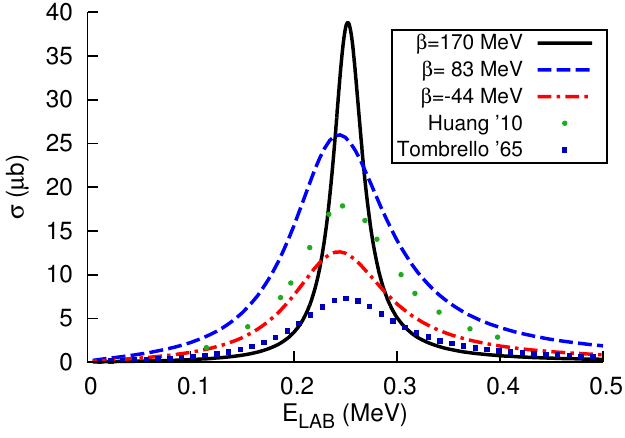} 
\end{center}
\caption{\protect M1 capture. Black curve with $\beta=170$ MeV, 
resonance energy and width fitted to experimental data. Blue dashed 
curve with $\beta=83$ MeV, red dot-dashed curve with 
$\beta=-44$ MeV, resonance energy and width fitted to Huang's 
potential model. Green circles are results from Huang et al., blue 
squares M1 capture using Tombrello's potential model in code CDXS+.}
\label{fig:M1capture}
\end{figure}

\begin{figure}[tbh]
\begin{center}
\includegraphics[width=0.48\textwidth,clip=true]{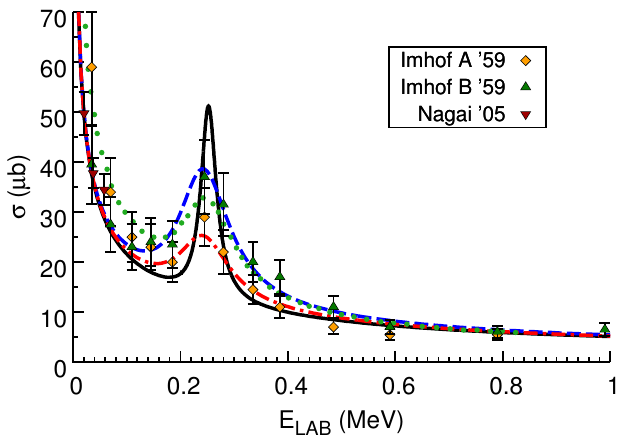}
\end{center}
\caption{\protect Capture cross section including E1 and M1 transition. We use 
$r_1^{(2^+)}=-1.47$ fm$^{-1}$, $r_1^{(1^+)} =-1.93$ fm$^{-1}$. Black solid 
curve with $\beta=170$ MeV and experimental resonance width, blue 
dashed curve $\beta=83$ MeV, red dot-dashed curve with 
$\beta=-44$ MeV, and Huang's potential model resonance width. Green 
dots are results from Huang et al. The experimental 
points are from Refs.~\cite{Imhof:1959,Nagai:2005,Blackmon:1996,Lynn:1991}.}
\label{fig:3plusCrossSection}
\end{figure}

All the elastic scattering parameters have been determined. 
The final expression in Eq.~(\ref{eq:M1sigma}) depends on two parameters 
$\beta^{(1)}$, $\beta^{(2)}$, that we fit to capture data 
near the resonance. 
The EFT couplings $(\Delta^{(3)}, h^{(3)})$ were matched to 
the position and width of the resonance in the $^5P_3$ elastic channel. 
The values of $\beta^{(i)}$s primarily affect the height of the cross 
section near the resonance, but not its position or width. 
If one follows the power counting naively then only the two-body currents 
contribute at LO, and the two $\beta^{(i)}$s are correlated. 
In the resummed amplitude, we find a similar behavior in our fits. 
Thus we use a common $\beta=\beta^{(1)}=\beta^{(2)}$. We find $\beta=170$ MeV 
when we fit to data set ImhofB~\cite{Imhof:1959} (more consistent with 
the low-energy $1/v$ behavior observed experimentally) 
using the experimental $3^+$ width. 
Instead, a fit to the same data set but with the $3^+$ width extracted from 
Huang's potential model phase shift provides $\beta=83$ MeV. 
Fitting to the data set  ImhofA with Huang's $3^+$ width give $\beta=-44$ MeV. 
The results are shown in Fig.~\ref{fig:M1capture}, 
~\ref{fig:3plusCrossSection}. Note that the authors assign a $20\%$ error 
to the data sets ImhofA and ImhofB~\cite{Imhof:1959}. 
This means away from the resonance where the cross section  is small, 
the errors are also small.  
This makes the region where the resonance contributes the most the least 
constraining in the fits. 

The fit with a wider resonance, $\Gamma_r=0.11$ MeV, seems to describe 
the data better than using the experimental width $\Gamma_r=0.031$ MeV, see Fig.~\ref{fig:3plusCrossSection}. 
Smaller errorbars near the resonance than the estimated $20\%$ ones 
would be useful to make stronger statements. 
Higher order EFT corrections estimated to be around $30\%$ (see 
Appendix~\ref{sec_powercounting}) could also make the fit better even with 
the narrower experimental width. 
With these considerations in mind, we see that with the current data 
the LO EFT does not reproduce the width seen in the capture. 
This could be an indication of the limitation of the single-particle 
approximation to describe the capture reaction near the resonance energy as 
it is towards the higher momenta region of the domain of applicability of 
the low energy EFT. 
Nonetheless, the same feature is observed in the microscopic calculation 
of Ref.~\cite{Bennaceur:1999hv}. 
The potential models get a wider width by coincidence 
and it is not in agreement with the experimental value $\Gamma_r=0.031$ MeV. 

To expand the range of applicability of the EFT to slightly 
higher momenta, one needs to include other missing degrees of freedom. 
The $\frac{1}{2}^{-}$ excited state of the $^7$Li core ($^7{\rm Li}^*$), 
which contributes only to the spin-1 channel, is the first to consider. 
It can be incorporated 
explicitly in the present halo EFT, since its energy remains close to the 
core ground state ($\sim 0.5$ MeV) and far from the first breakup channel 
($\sim 2.5$ MeV). In the $n$-$^7{\rm Li}$ c.m. system, the energy required 
to probe the $^7{\rm Li}^*$ involves momenta of $\sim 80$ MeV. 
In the present work, the higher incoming momentum considered is $\sim 40$ 
MeV, which justifies having the $^7{\rm Li}^*$ ``integrated out''. 
An analogous situation is the Delta resonance in chiral perturbation theory 
(the EFT for pions and nucleons) where the Delta can be integrated out 
of the theory when the energy relative to the pion-nucleon threshold is 
smaller than the nucleon-Delta mass 
splitting~\cite{Pascalutsa:2002pi,Long:2009wq,Cohen:2004kf}. 
Further, it can also be shown that the virtual contributions of $^7{\rm Li}^*$ 
to the ground state of $^8{\rm Li}$ in the neutron capture reaction is 
a subleading effect~\cite{HigaRupak}. 

Contributions from $d$-waves to the E1 transition, which are in principle 
suppressed by a factor of $(Q/\Lambda)^4$, may also become relevant with 
increasing energy~\cite{PhysRevC.68.045802}, depending on the desired 
accuracy. 
The next level of sophistication is the inclusion of alpha and 
triton degrees of freedom in a three-cluster treatment, which shall 
improve the description of the M1 capture reaction~\cite{Shul'gina1996197,
Grigorenko:1999mp,Descouvemont1994341,Descouvemont:2004hh}. 

\section{Conclusions}\label{sec_conclusions}

In the present work we extend the previous halo EFT calculation of the 
$^7{\rm Li}(n,\gamma)^8{\rm Li}$ capture reaction to include the 
complete E1 transition at leading order, as well as the leading M1 
capture at low energies. 
We present model-independent results that quantify the current 
uncertainties in nuclear theory in the single-particle approximation, 
therefore serving as a guide to its limitations and also highlighting 
where more precise experimental input is necessary for improvements. 

We include explicitly the E1 capture from $s$-waves to the excited state 
in $^8{\rm Li}$
that contributes about $10\%$ to the cross section at very low energy. 
The new EFT coupling constants associated to this process are completely 
determined by the binding energy of the excited state and the E1 thermal 
capture rate to the excited state~\cite{Lynn:1991}. 
For energies below $100$ keV, our results show the expected $1/v$ behavior 
also seen in potential models, however, differing by a sizable overall 
normalization, directly related to the effective range in the ground 
state channel $r_1^{(2^+)}$. 
This is the dominant source of uncertainty in this low energy region 
---it could be determined from accurate elastic scattering data and partial 
wave analysis, but due to the present lacking of the latter it is poorly 
known. A fit to E1 thermal capture data  gives 
$r_1^{(2^+)}\approx -1.47$ fm$^{-1}$, see also Ref.~\cite{Rupak:2011nk}. 

The M1 capture proceeds via the $3^+$ resonance near $E\sim 0.2$ MeV in 
the $^5P_3$ channel. 
It is suppressed at very low momentum due to the $p$-wave initial state. 
However, near the resonance there is an enhanced contribution 
to the total cross section that needs to be considered. In the halo EFT 
approach, we include and estimate the size of one-body 
(magnetic moment and orbital momentum couplings) and two-body currents 
that enters in the calculation. 
The one-body current contribution is consistent with the effective magnetic 
dipole operator used in potential models, but the two-body 
currents are new ingredients. These latter also renormalize loop 
contributions, and the corresponding couplings $\beta^{(1)}$ and 
$\beta^{(2)}$ are our free parameters to be determined from 
the capture data. The power counting for two-body currents was studied, 
and found to contribute at LO and N$^2$LO in the M1 and E1 capture, 
respectively. 

The available capture data near the resonance have large errors that 
hampers the quality of the fit. However, given the current data near the 
resonance, it seems that using the experimental 
resonance width $\Gamma_r=31$ keV in the $3^+$ elastic amplitude does a 
poor job in describing the M1 capture data. 
This was also observed in the microscopic calculation of 
Ref.~\cite{Bennaceur:1999hv}. A significantly larger width, 
about three times the experimental one, provides much better fits. 
This is roughly the width that one gets in potential models that are tuned 
to the resonance energy. 
This is a coincidence since, in principle, one should tune the potential 
models to reproduce not only the resonance energy $E_r$ but also 
the resonance width $\Gamma_r$ accurately. 
In the EFT formalism, the $p$-wave resonance requires two 
operators at LO that can be fixed by matching to the resonance energy and 
width. The fact that halo EFT is able to describe the resonance scattering 
(as shown in Section~\ref{sec_resonance}) but not to reproduce the M1 capture 
might indicate the limitation of the 
current approach. As discussed in Section~\ref{sec_crosssection}, the 
M1 capture could be on the outer edge of the range of applicability. 
To expand this EFT range, other degrees of freedom have to be incorporated. 
Within the present two-body treatment, the inclusion of the 
$\frac{1}{2}^{-}$ excited state in the $^7$Li core is the first step 
towards this goal. The next, more radical extension is to take the 
leading configuration of the $^7$Li as a bound state of elementary alpha 
and triton ``cores''. In such a three-body approach, not only the ground 
and the $\frac{1}{2}^{-}$ excited states in $^7$Li could be considered, 
but also the $\frac{7}{2}^{-}$ state which, according to microscopic 
approaches~\cite{Navratil:2010jn,Bennaceur:1999hv,Descouvemont:2004hh}, 
is important to properly describe the $3^+$ resonance. 
Nevertheless, the apparent discrepancy in the input related to 
the $3^+$ resonance width that is used in the $^5P_3$ elastic scattering 
and M1 capture reaction is unlikely to be resolved with current experimental 
information. More precise capture data around $0.22$ MeV (where 
the M1 capture dominates) is needed to conclusively state if the single-particle 
approximation is sufficient to describe the M1 capture in \nLi.

\acknowledgments
We thank C. Bertulani and S. Typel for discussions and help with 
potential model codes RADCAP~\cite{RADCAP} and  CDXS+~\cite{CDXSp}, 
respectively. We also thank T. Frederico, V. Guimar\~aes, H.-W. Hammer, 
D. R. Phillips, S. Typel, and U. van Kolck for valuable comments 
and discussions. 
This work is partially supported by 
the U.S. NSF Grant No. PHY-0969378 (LF and GR), 
the Dutch Stichting FOM under program 114 (RH) and HPC$^2$ Center for Computational Sciences at Mississippi State University (GR).

\appendix

\section{More on Power Counting}\label{sec_powercounting}

We have seen earlier that the contribution of the two-body currents 
relative to the loop diagrams in M1 capture scales as 
$L^{(i)}/(g_{[i]} Q h^{(3)} h^{(2)} )$, where $g_{[i]}$ is a combination of 
the gyromagnetic ratios $g_c$, $g_n$ that is of natural size in the $^5P_2$ 
channel but large in the $^3P_2$ channel. To estimate the size of 
two-body currents we also need estimates for $h^{(\eta)}$ 
that is related to the effective range. In this section we present the 
details of the power counting starting with the $p$-wave elastic channel. 
This also allows one to estimate the expansion parameter $Q/\Lambda$. 

In Sec.~\ref{sec_theory} we have fitted the ERE parameters $a_1^{(\eta)}$, 
$r_1^{(\eta)}$ in their respective $p$-channels. We found 
$\big|1/a_1^{(\eta)}\big|^{1/3}\sim 30-50$ MeV 
and $r_1^{(\eta)}/2\sim 100-250$ MeV. This is consistent with the situation in 
Ref.~\cite{Bedaque:2003wa}, where only $1/a_1^{(\eta)}$ 
is fine-tuned to scale as $\Lambda Q^2$ while $r_1^{(\eta)}\sim \Lambda$ 
obeys the naive expectation. 
Then in the EFT matching conditions, Eq.~(\ref{eq:a1r1}), 
\begin{align}
&-\frac{\pi}{[h^{(\eta)}]^2}= \frac{1}{2} r_1^{(\eta)}+\frac{3}{2}\lambda,
\nonumber\\
&-\frac{2\pi\mu\Delta^{(\eta)}}{[h^{(\eta)}]^2}=-1/a_1^{(\eta)}
+\frac{\pi}{2}\lambda^3,\nonumber
\end{align}
the renormalization scale $\lambda \sim Q$ from the loop momentum is a higher order 
contribution in a $Q/\Lambda$ expansion. 
We get 
$[h^{(\eta)}]^2= -2\pi/r_1^{(\eta)}\sim 1/\Lambda$ and 
$\mu \Delta^{(\eta)}= -1/(r_1^{(\eta)} a_1^{(\eta)})\sim Q^2$. 
We expand the dimer propagator as 
\begin{align}
D(p_0,\bm{p})=&D_{-2} (p_0,\bm{p})+ D_{-1}(p_0,\bm{p})+\cdots,
\end{align}
and the LO term is given by the bare propagator, 
\begin{align}
D_{-2}(p_0,\bm{p})=&\frac{1}{\Delta-\zeta^2/(2\mu)},\nonumber
\end{align}
since loop contributions enter at higher orders. 
The subscript indicates the scaling with powers of $Q$ and we have 
suppressed the superscript $\eta$ here and in the rest of this section. 
To renormalize the loop expansion systematically we write the couplings 
as~\cite{Chen:1999tn,Rupak:1999aa}
\begin{align}
h=&h_0+h_1+\cdots,\nonumber\\
\Delta=& \Delta_2+\Delta_3+\cdots,
\end{align}
stressing again that the subscripts bookkeep the powers in $Q$. 
Matching to the ERE we get 
\begin{align}
-\frac{\pi}{h_0^2}=\frac{1}{2} r_1, & & 2\pi\frac{h_1}{h_0^3}=
\frac{3}{2}\lambda,\nonumber\\
\Delta_2=-\frac{1}{\mu r_1 a_1}, & & \Delta_3= \frac{\pi}{2\mu r_1}
\lambda^3+\frac{3}{\mu r_1^2 a_1}\lambda. 
\label{eq:matchingA}
\end{align}
At next-to-leading order (NLO) the dimer propagator gets a contribution 
from the one-loop self-energy diagram, 
\begin{align}
D_{-1}(p_0,\bm{p})= &[iD_{-2}(p_0,\bm{p})]^2 \Sigma(p_0,\bm{p}),\nonumber\\
i\Sigma(p_0,\bm{p})= &i\frac{2h_0^2}{\mu}f_1(p_0,\bm{p})+i\Delta_3=
i\frac{h_0^2}{2\pi}\zeta^3-i\frac{3 h_0^2}{4\pi\mu}\lambda\zeta^2
+i\frac{h_0^2}{4\mu}\lambda^3+i\Delta_3 \nonumber\\
=&-i\frac{\zeta^3}{r_1}
+i\frac{3\zeta^2\lambda}{2\mu r_1}+\frac{3\lambda}{\mu r_1^2 a_1},
\end{align}
which is a $\lambda$-dependent result. However, 
the elastic amplitude in Eq.~(\ref{eq:EFT_Ap}),
\begin{align}
A(p)=&-h^2 \frac{p^2}{\mu^2}D({p^2}/{2\mu},0)\nonumber\\
\approx &-h_0^2 \frac{p^2}{\mu^2}D_{-2}({p^2}/{2\mu},0)
-h_0^2 \frac{p^2}{\mu^2}D_{-2}({p^2}/{2\mu},0)
\[D_{-1}({p^2}/{2\mu},0)/D_{-2}({p^2}/{2\mu},0)
+\frac{2h_1}{h_0} \] \nonumber \\
= &\frac{2\pi}{\mu}\frac{p^2}{-1/a_1+\frac{1}{2} r_1 p^2}
\left(1+\frac{ip^3}{-1/a_1+\frac{1}{2} r_1 p^2}\right),
\end{align}
is $\lambda$-independent as expected for a physical observable. 
This is in agreement with the expansion of the ERE amplitude up to NLO 
in $Q/\Lambda$. 
At LO the binding momentum is given by $\gamma=\sqrt{-2/(r_1 a_1)}$. 
The NLO term introduces a double pole which, at least formally, is 
suppressed in the $Q/\Lambda$ expansion. 
To treat the bound state consistently we rewrite the 
$p$-wave ERE expansion~\cite{Chen:1999tn,Rupak:1999aa} as: 
\begin{align}
p^3\cot\delta(p)=&\gamma^3 +\frac{1}{2} s (p^2+\gamma^2)+\cdots,\nonumber\\
-1/a_1=&\gamma^3+\frac{1}{2}s\gamma^2+\cdots, \nonumber\\
r_1=&s+\cdots,
\end{align}
and define the EFT couplings as 
\begin{align}\label{eq:matchingB} 
-\frac{\pi}{h_0^2}=\frac{1}{2} r_1, & & 2\pi\frac{h_1}{h_0^3}=
\frac{3}{2}\lambda,\nonumber\\
\Delta_2=\frac{\gamma^2}{2\mu }, & & \Delta_3=
\frac{\pi}{2\mu r_1}\lambda^3-\frac{3\gamma^2}{2\mu r_1}\lambda
+\frac{\gamma^3}{\mu r_1}. 
\end{align}
The amplitude is now written as 
\begin{align}
A(p)=\frac{2\pi}{\mu}\frac{2p^2}{r_1(\gamma+ip)(\gamma-ip)}
\[1+ 2\frac{p^2+ip\gamma-\gamma^2}{r_1(\gamma-ip)}\],
\end{align}
where the NLO correction contributes a factor of $-3\gamma/r_1$ to the 
residue at the pole $p=i\gamma$ without introducing any spurious double 
pole. This correction to the residue at the pole is consistent with the 
wave function renormalization calculated earlier, 
\begin{align}
h^2 {\cal Z}= -\frac{2\pi}{3\gamma+r_1}.
\end{align}
We keep the complete result instead of expanding. A convenient approach 
to recover the complete result at NLO without resumation is to use the so 
called ``zed"-parameterization~\cite{Phillips:1999hh}. In this approach 
one would define, for example, $h_0^2= -2\pi({\cal Z}-1)/(3\gamma)$, 
$h_1=3h_0^3\lambda/(4\pi)$ and recover, for the wave function renormalization, 
\begin{align}
{\cal Z}=1+({\cal Z}-1)+0+0+\cdots,
\end{align}
where ${\cal Z}-1\sim Q$ is treated as a perturbation, see 
Ref.~\cite{Phillips:1999hh} for details. 
Here we will simply use the resummed result for 
the $2^+$ bound state and the $1^+$ excited state. 

So far we have discussed the fine-tuning required to reproduce a shallow 
($p$-wave) bound or virtual state. The power counting above also applies 
to low lying resonances. However, for these cases there is an 
additional fine-tuning that is purely kinematical that was discussed in 
Refs.~\cite{Bedaque:2003wa,Pascalutsa:2002pi}. 
This second fine-tuning requires the loop 
contribution to be treated non-perturbatively at energies near the 
resonance. Since we consider energies near the resonance in our capture 
calculations, we resum the loop contributions in the $3^+$ initial state. 

Next we come back to the role of two-body currents 
$L/(g_{[i]} Q h^2 )$ in the M1 capture. 
Since $h^2\sim 1/\Lambda$ as shown above, the relative contribution for a 
natural two-body current scales as $\Lambda/(g_{[i]} Q)$. 
At LO and from the specific numerical values of $g_{[i]}$, 
one notices that in the $^5P_2$ channel only the two-body current enters, 
while in the $^3P_2$ case both two-body and 
one-body (magnetic moment) currents contribute. 
In principle, for a systematic treatment one could write 
$\beta^{(2)}=\beta^{(2)}_0+\beta^{(2)}_1+\cdots$ and perform the 
perturbative renormalization outlined above for the $^5P_2$ channel, 
while keeping the full loop contribution in the $^3P_2$ channel. 
We verified that such a treatment satisfies the power counting estimates. 
As mentioned in Sec.~\ref{sec_crosssection}, 
we keep the loop contribution at LO in both $^3P_2$ and $^5P_2$ channels 
for convenience. 

The scaling of two-body currents that appear in the E1 case is different 
than in the M1 capture.  To keep the discussion fairly general, let us introduce a dimer 
field $\pi^{(s\mathrm{-wave})}$ for the two initial state $s$-wave channels $^5S_2$, $^3S_1$. 
Then the relative contribution of the two-body to one-body current in the E1 capture  
is generically $L_\mathrm{E1}k_0\mu h^{(s\mathrm{-wave})}/ [h \Delta^{(s\mathrm{-wave})}]$, where 
we  considered the operator 
$e L_\mathrm{E1} \phi^\dagger_{i j} E_x  \pi_{y z} T_{x y z i j}$ for transition 
from $^5S_2$ to $^5P_2$ ground state, for illustration.  $h^{(s\mathrm{-wave})}\sim h$ is the $\pi$-nucleon-core coupling and 
$\Delta^{(s\mathrm{-wave})}$ the dimer propagator in the $s$-wave.  In the power counting one either takes 
$\Delta^{(s\mathrm{-wave})}\sim \Lambda$ to treat $s$-wave interaction as perturbative as would be the case for small natural sized scattering length (the $^3S_1$ channel for momenta $p\lesssim 227$ MeV) or take $\Delta^{(s\mathrm{-wave})}\sim Q$ to treat  
 $s$-wave interaction as non-perturbative as would be the case for large  unnatural sized scattering length 
 (the $^5S_2$ channel around $p\sim 54$ MeV).   
For a natural $L_\mathrm{E1}$ given by dimensional analysis, 
the relative contribution $L_\mathrm{E1}k_0\mu h^{(s\mathrm{-wave})}/ [h \Delta^{(s\mathrm{-wave})}]$ scales as either $Q^2$ for perturbative 
or  as $Q$ for non-perturbative $s$-wave interaction  in the initial state.  The former is a N$^2$LO contribution whereas the latter is a NLO contribution.

\section{Normalization of states}\label{sec_normalization}

In this work we adopt the following definitions for the nucleon and core 
field operators 
\begin{eqnarray}
\psi_{N}({\mbox{\boldmath $x$}})&=&
\int\frac{d^3p}{(2\pi)^3}\frac{1}{\sqrt{2M_{N}}}
\sum_{s}N_{s}({\mbox{\boldmath $p$}})U_{N}^{(s)}({\mbox{\boldmath $p$}})
e^{i{\mbox{\scriptsize\boldmath $p$}}\cdot{\mbox{\scriptsize\boldmath $x$}}}\,,
\nonumber\\&&
\sum_{s}U_{N}^{(s)}({\mbox{\boldmath $p$}})
U_{N}^{(s)\,\dagger}({\mbox{\boldmath $p$}})=
2M_{N}\sum_{s}\chi^{(s)}\chi^{(s)\,\dagger}=2M_{N}\,I_{2\times 2}\,,
\\[1mm]
\psi_{C}({\mbox{\boldmath $x$}})&=&
\int\frac{d^3p}{(2\pi)^3}\frac{1}{\sqrt{2M_{C}}}
\sum_{r}C_{r}({\mbox{\boldmath $p$}})U_{C}^{(r)}({\mbox{\boldmath $p$}})
e^{i{\mbox{\scriptsize\boldmath $p$}}\cdot{\mbox{\scriptsize\boldmath $x$}}}\,,
\nonumber\\&&
\sum_{r}U_{C}^{(r)}({\mbox{\boldmath $p$}})
U_{C}^{(r)\,\dagger}({\mbox{\boldmath $p$}})=
2M_{C}\sum_{r}\xi^{(r)}\xi^{(r)\,\dagger}=2M_{C}\,I_{4\times 4}\,,
\\[1mm]
&&\{\psi_{N,a}({\mbox{\boldmath $x$}}),
\psi^{\dagger}_{N,b}({\mbox{\boldmath $y$}})\}
=\{\psi_{C,a}({\mbox{\boldmath $x$}}),
\psi^{\dagger}_{C,b}({\mbox{\boldmath $y$}})\}
=\delta^{(3)}({\mbox{\boldmath $x$}}-{\mbox{\boldmath $y$}})\delta_{ab}\,,
\nonumber \\[1mm]
&&
\{N_{a}({\mbox{\boldmath $p$}}),N^{\dagger}_{b}({\mbox{\boldmath $q$}})\}
=\{C_{a}({\mbox{\boldmath $p$}}),C^{\dagger}_{b}({\mbox{\boldmath $q$}})\}
=(2\pi)^3\delta^{(3)}({\mbox{\boldmath $p$}}-{\mbox{\boldmath $q$}})
\delta_{ab}\,.\nonumber
\end{eqnarray}
$\chi^{(s)}$ and $\xi^{(r)}$ are nucleon and core spinors in the fundamental
representations of spins $1/2$ and $3/2$, respectively.
One-neutron states are defined as
\begin{eqnarray}
&&|{\mbox{\boldmath $p$}},s\rangle=\sqrt{2M_{N}}\,
N_{s}^{\dagger}({\mbox{\boldmath $p$}})
|0\rangle\,,\qquad
\Rightarrow\qquad \langle{\mbox{\boldmath $p$}},s|
{\mbox{\boldmath $q$}},s'\rangle=2M_{N}\,(2\pi)^3
\delta^{(3)}({\mbox{\boldmath $p$}}-{\mbox{\boldmath $q$}})\delta_{ss'}\,,
\nonumber\\[2mm]&&
\psi_{N}(x)|{\mbox{\boldmath $p$}},s\rangle=\int\frac{d^3q}{(2\pi)^3}
\frac{1}{\sqrt{2M_{N}}}\sum_{s'}
U_{N}^{(s')}({\mbox{\boldmath $q$}})e^{i{\mbox{\scriptsize\boldmath $q$}}\cdot
{\mbox{\scriptsize\boldmath $x$}}}N_{s'}({\mbox{\boldmath $q$}})
\,\sqrt{2M_{N}}\,N_{s}^{\dagger}({\mbox{\boldmath $p$}})|0\rangle
\nonumber\\[0mm]&&\hspace{1.65cm}
=\int\frac{d^3q}{(2\pi)^3}\sum_{s'}
U_{N}^{(s')}({\mbox{\boldmath $q$}})e^{i{\mbox{\scriptsize\boldmath $q$}}\cdot
{\mbox{\scriptsize\boldmath $x$}}}\{N_{s'}({\mbox{\boldmath $q$}}),
N_{s}^{\dagger}({\mbox{\boldmath $p$}})\}|0\rangle
=e^{i{\mbox{\scriptsize\boldmath $p$}}\cdot{\mbox{\scriptsize\boldmath $x$}}}
U_{N}^{(s)}({\mbox{\boldmath $p$}})|0\rangle\,.
\end{eqnarray}
Analogously for one-core states. Generalization to multi-particle states
is straightforward.

\subsection{Projection operators}
For each partial wave we construct the corresponding projection operators 
from the relative core-nucleon velocity, the spin-1/2 
Pauli matrices $\sigma_{i}$'s, and the following spin-1/2 to spin-3/2 
transition matrices 
\begin{align}
S_1=&\frac{1}{\sqrt{6}}\(\begin{array}{cccc}
-\sqrt{3} & 0 & 1 & 0\\
0&-1&0&\sqrt{3}
\end{array}\) , \,\,
&{} 
&S_2 = -\frac{i}{\sqrt{6}} \(\begin{array}{cccc}
\sqrt{3} & 0 & 1 & 0\\
0&1&0&\sqrt{3}
\end{array}\) , \,\,
S_3 = \frac{2}{\sqrt{6}} \(\begin{array}{cccc}
0 & 1 & 0 & 0\\
0&0&1&0
\end{array}\),
\end{align}
which satisfy
\begin{equation}
S_{i}S^{\dagger}_{j}=\frac{2}{3}\delta_{ij}-\frac{i}{3}\epsilon_{ijk}
\sigma_{k}\,,\qquad
S_{i}^{\dagger}S_{j}=\frac{3}{4}\delta_{ij}-\frac{1}{6}\big\{J_{i}^{(3/2)},
J_{j}^{(3/2)}\big\}+\frac{i}{3}\epsilon_{ijk}J_{k}^{(3/2)}\,,
\end{equation}
where $J_{i}^{(3/2)}$'s are the generators of the spin-3/2. 
We construct the Clebsch-Gordan coefficient matrices 
\begin{align}
F_i=&-\frac{i\sqrt{3}}{2}\sigma_2 S_i, 
&{}
&Q_{i j} = -\frac{i}{\sqrt{8}}\sigma_2\big(\sigma_i S_i+\sigma_j S_i\big),
\label{eq:ncorespinmatrices}
\end{align}
and define the tensors 
\begin{align}
R_{ijxy}&=\frac{1}{2}\left(\delta_{i x}\delta_{j y}+\delta_{i y}\delta_{j x} 
-\frac{2}{3}\delta_{i j}\delta_{x y}\right),
\nonumber\\
T_{xyz i j}&=\frac{1}{2}\Big(\epsilon_{x z i}\delta_{y j}
+\epsilon_{x z j}\delta_{y i}
+\epsilon_{y z i}\delta_{x j}+\epsilon_{y z j}\delta_{x i} \Big),
\nonumber\\
G_{i j k l m q}&= \frac{1}{6}\bigg\{ -\frac{2}{5}\Big[
\delta_{q m} (\delta_{i l}\delta_{j k}+\delta_{j l}\delta_{i k}
+\delta_{i j}\delta_{k l}) +\delta_{q l} (\delta_{i m}\delta_{j k}
+\delta_{j m}\delta_{i k}+\delta_{i j}\delta_{k m}) 
\nonumber\\
&+\delta_{l m} (\delta_{i q}\delta_{j k}+\delta_{j q}\delta_{i k}
+\delta_{i j}\delta_{k q})\Big] 
+(\delta_{i l}\delta_{j m}\delta_{k q} +\delta_{i l}\delta_{j q}\delta_{k m}) 
\nonumber\\
&(\delta_{j l}\delta_{k m}\delta_{i q}+\delta_{j l}\delta_{k q}\delta_{i m}) 
+(\delta_{k l}\delta_{i m}\delta_{j q}+\delta_{kl}\delta_{i q}\delta_{j m})
\bigg\} ,
\label{eq:RTGtensorsdef}
\end{align}
that assure the correct number of independent indices for a given total 
angular momentum. The latter have the following properties, 
\begin{eqnarray}
R_{ijlm}&=&R_{jilm}=R_{ijml}=R_{lmij}\,, \nonumber
\\[1mm]
T_{ijklm}&=&T_{jiklm}=T_{ijkml}=-T_{lmkij}\nonumber \,,
\\[1mm]
G_{ijklmq}&=&G_{jiklmq}=G_{kjilmq}=G_{ikjlmq}=G_{ijkmlq}=G_{ijkqml}
=G_{ijklqm}=G_{lmqijk}\,,
\nonumber \\[4mm]
&\Rightarrow&
R_{ijxy}R_{xylm}=R_{ijlm}\,,\qquad
R_{ijxy}T_{xyklm}=T_{ijkxy}R_{xylm}=T_{ijklm}\,,\nonumber\\[0.0mm]&&
G_{abcijk}G_{ijklmn}=G_{abclmn}=R_{abxy}G_{xyclmn}\,.
\end{eqnarray}
We introduce the photon vector ($\varepsilon^{(\gamma)}_{i}$), 
spin-1 ($\varepsilon_{j}$), spin-2 ($\varepsilon_{ij}$) 
and spin-3 tensor ($\varepsilon_{ijk}$) polarizations, obeying the 
following sums~\cite{Choi:1992,*Fleming:1999ee}, 
\begin{equation}
\sum_{\rm pol.}\varepsilon^{(\gamma)}_{i}\varepsilon^{(\gamma)*}_{j}=
\delta_{ij}-\frac{k_ik_j}{k^2}\,,
\quad
\sum_{\rm pol.avg.}\varepsilon_{i}\varepsilon^{*}_{j}=\frac{\delta_{ij}}{3}\,,
\quad
\sum_{\rm pol.avg.}\varepsilon_{ij}\varepsilon^{*}_{lm}=\frac{R_{ijlm}}{5}\,,
\quad
\sum_{\rm pol.avg.}\varepsilon_{ijk}\varepsilon^{*}_{lmq}
=\frac{G_{ijklmq}}{7}\,.
\label{eq:poltensordefs}
\end{equation}
All these elements, together with the matrices $F_i$ and $Q_{ij}$ from 
Eq.~(\ref{eq:ncorespinmatrices}), allow us to build the operators, 
in coordinate space, 
\begin{align}
&P^{(3P_1)}_i=\sqrt{\frac{3}{2} }F_x\P_y \epsilon_{i x y},\nonumber\\
&P^{(^3P_2)}_{i j}=\sqrt{3} F_x \P_y\ R_{x y i j },\nonumber\\
&P^{(^5P_1)}_i=\sqrt{\frac{9}{5}} Q_{i x}\P_ x ,\nonumber\\
&P^{(^5P_2)}_{i j}=\frac{1}{\sqrt{2}} Q_{x y} \P_z\ T_{x y z i j},\nonumber\\
&P^{(^5P_3)}_{i j k}=\sqrt{3} Q_{x y}\P_z G_{i j k x y z},
\label{eq:projdefr-space}
\end{align}
or in the momentum space, 
\begin{eqnarray}
\hat P^{({}^{3}S_{1})}_{j}&=&F_{j}\,,
\label{eq:3S1projj}\nonumber
\\[1mm]
\hat P^{({}^{5}S_{2})}_{ij}&=&Q_{ij}\,,
\label{eq:5S2projij}\nonumber
\\[1mm]
\tilde P^{({}^{3}P_{1})}_{j}&=&
\bar p_{z}\,i\sqrt{\frac{3}{2}}\,F_{y}\epsilon_{jyz}
=\bar p\,i\sqrt{\frac{3}{2}}\,F_{y}\,\frac{\bar p_{z}}{\bar p}\,\epsilon_{jyz}
=\bar p\,\hat P^{({}^{3}P_{1})}_{j}\,,
\label{eq:3P1projj}\nonumber
\\[1mm]
\tilde P^{({}^{5}P_{1})}_{j}&=&
\bar p_{z}\,i\frac{3}{\sqrt{5}}\,Q_{jz}
=\bar p\,i\frac{3}{\sqrt{5}}\,\frac{\bar p_{z}}{\bar p}\,Q_{jz}
=\bar p\,\hat P^{({}^{5}P_{1})}_{j}\,,
\label{eq:5P1projj}\nonumber
\\[1mm]
\tilde P^{({}^{3}P_{2})}_{ij}&=&
\bar p_{z}\,i\sqrt{3}F_{y}R_{yzij}
=\bar p\,i\sqrt{3}F_{y}\,\frac{\bar p_{z}}{\bar p}\,R_{yzij}
=\bar p\,\hat P^{({}^{3}P_{2})}_{ij}\,,
\label{eq:3P2projij}\nonumber
\\[1mm]
\tilde P^{({}^{5}P_{2})}_{ij}&=&
\bar p_{z}\,\frac{i}{\sqrt{2}}\,Q_{xy}T_{xyzij}
=\bar p\,\frac{i}{\sqrt{2}}\,Q_{xy}\,\frac{\bar p_{z}}{\bar p}\,T_{xyzij}
=\bar p\,\hat P^{({}^{5}P_{2})}_{ij}\,,
\label{eq:5P2projij}\nonumber
\\[1mm]
\tilde P^{({}^{5}P_{3})}_{ijk}&=&
\bar p_{z}\,i\sqrt{3}\,Q_{xy}G_{xyzijk}
=\bar p\,i\sqrt{3}\,Q_{xy}\,\frac{\bar p_{z}}{\bar p}\,G_{xyzijk}
=\bar p\,\hat P^{({}^{5}P_{3})}_{ijk}\,,
\label{eq:5P3projijk}
\end{eqnarray}
where we used 
\begin{equation}
\mu=\frac{M_{C}M_{N}}{M_{C}+M_{N}}\,,
\qquad r=\frac{M_{C}-M_{N}}{M_{C}+M_{N}}\,,
\qquad\qquad 
\bar p=p^{-}-rp^{+}=\frac{p_{c}-p_{n}}{2}-r\,\frac{p_{c}+p_{n}}{2}\,.
\end{equation}
The projector operators, defined by 
\begin{equation}
{\cal P}^{(\eta)}=\hat P^{(\eta)}_{[i]}\varepsilon_{[i]}
\end{equation}
where $[i]$ representing collectively the indices in a given channel, 
satisfy 
\begin{equation}
\sum_{\rm pol.avg.}{\rm Tr}\big[{\cal P}^{(\eta)}
{\cal P}^{(\eta)\dagger}\big]=1\,.
\label{eq:polavgprojectors}
\end{equation}
This can be straightforwardly verified from the relations above and 
\begin{equation}
{\rm Tr}[F_iF^{\dagger}_j]=\delta_{ij}\,,\qquad
{\rm Tr}[Q_{ij}Q^{\dagger}_{lm}]=R_{ijlm}\,. 
\end{equation}

\bibliographystyle{apsrev4-1}
\bibliography{Reference.bib}

\end{document}